\documentclass[a4paper]{article}

\usepackage[english]{babel}
\usepackage[utf8]{inputenc}
\usepackage[T1]{fontenc}

\usepackage{geometry}
\usepackage{setspace}
\usepackage{csquotes}
\usepackage{etoolbox}
\usepackage{amsmath}
\usepackage{tikz}
\usepackage{amssymb}
\usepackage{bm}
\usepackage{bbm}
\usepackage{graphicx}
\usepackage[colorinlistoftodos]{todonotes}
\usepackage[colorlinks=true, allcolors=blue]{hyperref}
\usepackage{xcolor}
\usepackage{float}
\usepackage{comment}
\usepackage{fancyvrb}
\usepackage{booktabs}
\usepackage{multirow}
\usepackage{pdflscape}
\usepackage{afterpage} 
\usepackage{caption}
\usepackage{multicol}
\usepackage{subcaption}
\usepackage{adjustbox}
\usepackage{threeparttable}
\usepackage{pdflscape}
\usepackage{longtable}
\usepackage{sansmath}
\usepackage{outlines}
\usepackage[colorinlistoftodos]{todonotes}
\usepackage{rotating} 
\usepackage{array}
\usepackage{ragged2e} 
\usepackage[backend=biber, sorting = none, citestyle = numeric, bibstyle = apa, doi=false,isbn=false,url=false,eprint=false, apamaxprtauth=99]{biblatex}

\makeatletter
\RequireBibliographyStyle{numeric}
\makeatother

\newcolumntype{R}[1]{>{\RaggedRight}p{#1}}
\usepackage[title]{appendix}	

\usepackage{indentfirst}

\usepackage[hang,flushmargin,perpage]{footmisc} 

\usepackage{tablefootnote}

\usepackage{tikz}
\usetikzlibrary{positioning}
\usepackage{adjustbox}
\RequirePackage{import}
\usetikzlibrary{decorations.pathreplacing}

\definecolor{darkcyan}{rgb}{0.0, 0.55, 0.55}
\definecolor{darkpastelred}{rgb}{0.76, 0.23, 0.13}
\definecolor{orangepeel}{rgb}{1.0, 0.62, 0.0}

\usepackage{enumitem}

\usepackage{wrapfig}

\title{When exposure affects subgroup membership: Framing relevant causal questions in perinatal epidemiology and beyond }
\author{Shalika Gupta$^1$ \and Laura B. Balzer$^2$ \and Moses R. Kamya$^3$ \and Diane V. Havlir$^4$ \and Maya L. Petersen$^{1,2,5}$}
\date{%
    \normalsize
    $^1$Division of Epidemiology, University of California, Berkeley\\%
    $^2$Division of Biostatistics, University of California, Berkeley\\%
    $^3$Department of Medicine, Makerere University College of Health Sciences\\%
    $^4$Division of HIV, Infectious Diseases, and Global Medicine, Department of Medicine, University of California, San Francisco\\%
    $^5$UCSF-UC Berkeley Program in Computational Precision Health\\%
    }

\begin{document}

\maketitle

\begin{abstract}
 Perinatal epidemiology often aims to evaluate exposures on infant outcomes. When the exposure affects the composition of people who give birth to live infants (e.g., by affecting fertility, behavior, or birth outcomes), this ``live birth process'' mediates the exposure effect on infant outcomes. Causal estimands previously proposed for this setting include the total exposure effect on composite birth and infant outcomes, controlled direct effects (e.g., enforcing birth), and principal stratum direct effects. Using perinatal HIV transmission in the SEARCH Study as a motivating example, we present two alternative causal estimands: 1) conditional total effects; and 2) conditional stochastic direct effects, formulated under a hypothetical intervention to draw mediator values from some distribution (possibly conditional on covariates). The proposed conditional total effect includes impacts of an intervention that operate by changing the types of people who have a live birth and the timing of births. The proposed conditional stochastic direct effects isolate the effect of an exposure on infant outcomes excluding any impacts through this live birth process. In SEARCH, this approach quantifies the impact of a universal testing and treatment intervention on infant HIV-free survival absent any effect of the intervention on the live birth process, within a clearly defined target population of women of reproductive age with HIV at study baseline. Our approach has implications for the evaluation of intervention effects in perinatal epidemiology broadly, and whenever causal effects within a subgroup are of interest and exposure affects membership in the subgroup.
\end{abstract}


\newpage
\section{Introduction}
\begin{figure}[htb]
	\centering
	\begin{tikzpicture}

\node[text centered, align=center] (A) {Exposure};
\node[right = .5 of A, text centered, align = center] (Z) {Pregnancy};
\node[right= .5 of Z, text centered, align = center] (Y) {Infant outcome};

\draw[->, line width= 0.5] (A) -- node[above,font=\footnotesize]{}  (Z);
\draw [->, line width= 0.5] (Z) -- node[above,font=\footnotesize]{}  (Y);
\draw[->, line width=0.5] (A) to  [out=270,in=270, looseness=0.9] node[below, font=\footnotesize]{} (Y);

\end{tikzpicture}
	\caption{Directed Acyclic Graph (DAG) with pregnancy as a mediator of the effect of an exposure on infant outcomes. For illustration, we focus on the setting of a randomized trial and omitting unmeasured factors.}
	\label{dag:pregnancy}
	\end{figure}
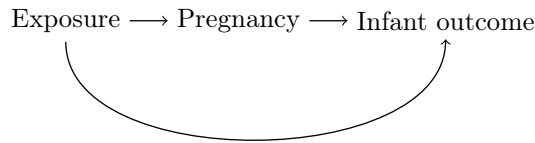

Perinatal epidemiology often evaluates the effects of exposures on infant outcomes when the exposure precedes or occurs during pregnancy and may affect the probability of live birth (Figure \ref{dag:pregnancy}). This poses a profound challenge to specifying relevant and interpretable causal effects of exposures on infant outcomes. Among individuals at risk of pregnancy, infant outcomes are only defined once a live birth occurs. Importantly, the timing of birth and the risk of adverse infant outcomes due to baseline characteristics of the individuals who give birth may differ by exposure status, even when the exposure is randomized. This fundamental challenge, although particularly acute in perinatal epidemiology, occurs whenever causal effects are of interest within a subset of the target population, and membership in that subset is affected by the exposure.\supercite{hernan_hazards_2010}

In perinatal epidemiology, Chiu et al.\supercite{chiu_effect_2020} proposed defining total effects based on a composite endpoint of live birth and infant outcomes. This definition implicitly assumes that all individuals in the population are pregnancy seeking (or all are not pregnancy seeking), which may be applicable in some settings, but not all.\supercite{chiu_effect_2020} It further includes any exposure effects on the live birth process, which may or may not be of interest. 
A range of mediation-based causal estimands have been proposed in perinatal epidemiology to disentangle these effects, including controlled direct, separable, and principal stratum direct effects.\supercite{chiu_effect_2020, young_identified_2021, stensrud_conditional_2023} While useful in some settings, each has limitations.\supercite{young_identified_2021, snowden_conceiving_2020}  
	
	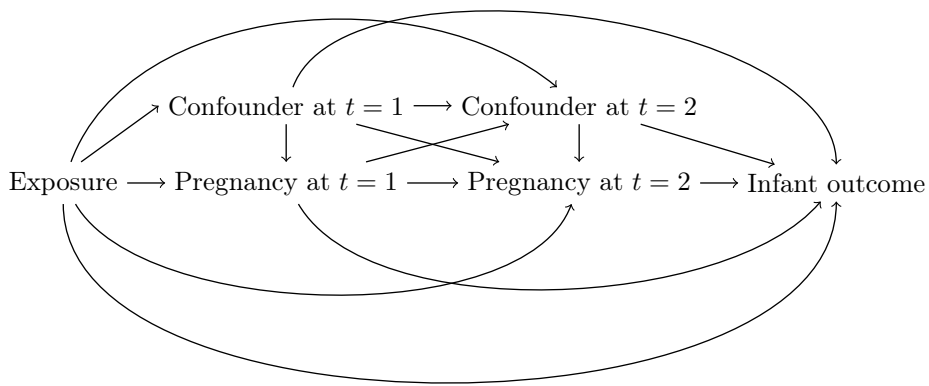
\begin{figure}[htb]
	\centering
	\begin{tikzpicture}

\node[text centered, align = center] (L1) {Confounder at $t=1$};
\node[right = .5 of L1, text centered, align = center] (L2) {Confounder at $t=2$};
\node[below = .5 of L1, text centered, align = center] (Z1) {Pregnancy at $t=1$};
\node[left = 0.5 of Z1, text centered, align=center] (A) {Exposure};
\node[below = .5 of L2, text centered, align = center] (Z2) {Pregnancy at $t=2$};
\node[right= .5 of Z2, text centered, align = center] (Y) {Infant outcome};

\draw[->, line width = 0.5] (A) to [out=50, in=180, looseness=0] node[left,font=\footnotesize]{}  (L1);
\draw[->, line width = 0.5] (A) to [out=70, in=140, looseness=0.9] node[above,font=\footnotesize]{}  (L2);
\draw[->, line width = 0.5] (A) -- node[above,font=\footnotesize]{}  (Z1);
\draw[->, line width = 0.5] (A) to [out=300, in=250, looseness=0.7] node[above,font=\footnotesize]{}  (Z2);
\draw[->, line width = 0.5] (A) to  [out=270,in=270, looseness=0.8] node[below, font=\footnotesize]{} (Y);

\draw [->, line width = 0.5] (L1) -- node[above,font=\footnotesize]{}  (L2);
\draw [->, line width = 0.5] (L1) -- node[above,font=\footnotesize]{}  (Z1);
\draw [->, line width = 0.5] (L1) -- node[above,font=\footnotesize]{}  (Z2);
\draw [->, line width = 0.5] (L1) to [out=70, in=90, looseness=0.7] node[above,font=\footnotesize]{}  (Y);

\draw [->, line width = 0.5] (Z1) -- node[above,font=\footnotesize]{}  (L2);
\draw [->, line width = 0.5] (Z1) -- node[above,font=\footnotesize]{}  (Z2);
\draw [->, line width = 0.5] (Z1) to [out=300, in=230,looseness=0.7] node[above,font=\footnotesize]{}  (Y);

\draw [->, line width = 0.5] (L2) -- node[above,font=\footnotesize]{}  (Z2);
\draw [->, line width = 0.5] (L2) -- node[above,font=\footnotesize]{}  (Y);

\draw [->, line width = 0.5] (Z2) -- node[above,font=\footnotesize]{}  (Y);
\end{tikzpicture}
	\caption{DAG with pregnancy as a mediator of a randomized exposure on infant outcomes including time-varying confounders at two time points (and omitting unmeasured factors).}
	\label{dag:pregnancy2}
	\end{figure}

In this paper, we first discuss an alternative definition of total effects in perinatal studies, using a composite outcome as in Chiu et al.,\supercite{chiu_effect_2020} but instead focusing on a conditional total effect (CTE) among counterfactual live births. The proposed conditional total effect includes by definition any impact of an intervention that operates by changing the types of people who give birth and timing of live births. Such differences are not a failure of causal identification; instead, these differences are a result of the paths by which an exposure can affect infant outcomes in settings where group membership may be affected by treatment.

Next, in order to isolate the exposure effect on infant outcomes not mediated by the birth process, we present an alternative class of conditional mediation causal parameters that are identifiable, interpretable, and which answer distinct questions in perinatal epidemiology from those previously proposed. Following the causal roadmap,\supercite{petersen_causal_2014} we consider a longitudinal data structure with time-varying covariates and a time-varying mediator (live birth process) (Figure \ref{dag:pregnancy2}). We present conditional stochastic direct effect (CSDE) parameters formulated using stochastic interventions on the live birth process, which allows us to isolate the exposure effect on infant outcomes \textit{not} due to any effects on the types of people who give birth and timing of live births (i.e., within the same counterfactual population at-risk), and discuss identification assumptions for these parameters. The SEARCH Study provides a detailed illustration. Our approach is applicable to a wide range of perinatal epidemiologic research, as well as more generally to the broad class of studies in which the causal effect of interest is defined among a subset of the population, and membership in this subset may be affected by the exposure.

\subsection{The SEARCH Study}
The SEARCH Study was a cluster randomized trial that evaluated the effectiveness of  universal antiretroviral treatment (ART) and person-centered HIV care on population health in rural Kenya and Uganda.\supercite{havlir_hiv_2019, ayieko_patient-centered_2019} The study tested 90\% of adults for HIV at baseline\supercite{chamie_hybrid_2016} and ascertained vital status and HIV status after three years for infants born during the study. In pre-specified analyses, the SEARCH intervention reduced perinatal HIV transmission and increased infant HIV-free survival among women of reproductive age living with HIV.\supercite{gupta_population-level_2020} However, the extent to which this total effect may have operated through intervention-induced changes to the counterfactual population of infants at-risk, due to timing of births (in the context of time-varying transmission risk across the pre and postnatal risk periods) 
and/or the types of people who gave birth (e.g., through changes to maternal survival, fertility, or behavior) remains unclear. CSDEs provide an approach to answering this question. 

\section{Previous Mediation-Based Approaches}
A number of mediation-based perinatal epidemiology parameters have been previously proposed. In particular, prior authors\supercite{chiu_effect_2020, young_identified_2021} have sought to exclude from the total effect of the exposure on infant outcomes any impact of the exposure on the
live birth process (the mediator) through: 1) hypothetical static interventions on the live birth process (i.e.,
controlled direct effects); 2) restriction to the group of individuals who would have had a live birth irrespective of
exposure (i.e., principal stratum direct effects); and, 3) partitioning of the exposure into distinct
components that operate in different ways on the live birth process and infant outcomes (i.e., separable direct
effects). 

Each of these approaches has limitations. First, a controlled direct effect evaluates the exposure effect on infant outcomes if all individuals in the target population were to give birth. This type of hypothetical intervention may be both hard to interpret and poorly supported by the data, particularly if the probability of live birth is low and/or many are not seeking pregnancy. A controlled direct effect quantifies the impact of the SEARCH intervention on infant HIV-free survival if all reproductive age women living with HIV at study baseline were to have a birth during the follow-up period, a counterfactual scenario that is neither realistic nor informative (Figure \ref{dag:interventions}\subref{dag:cde}). 

Second, identification of principal stratum direct effects requires strong assumptions, such as an assumption that no `defiers' exist. In a perinatal setting with live birth as a mediator, this corresponds to assuming that there are no individuals who would have a live birth if unexposed, but not if exposed. Moreover, if exposure affects the timing of births within the principal stratum, this direct effect parameter would still include any impact that operates through time at risk. For example, in SEARCH this would include any effect due to changes in timing of birth during follow-up and corresponding shifts in the distribution of infant exposure to pre versus and postnatal transmission risk. 

Third, separable direct effect parameters are predicated on the decomposition of the exposure into two components, one that acts directly on the outcome without affecting the mediator (live birth process) and a second that operates only on the (infant) outcome through the mediator. Often, this decomposition, which requires specifying mechanisms by which exposure affects infant survival but not maternal survival, may not be feasible. In the SEARCH trial, where the outcome of interest involves infant HIV-free survival and the mediator includes maternal survival, we cannot disentangle the portion of the intervention that impacts only the outcome without affecting the mediator.

To answer many questions in perinatal epidemiology, alternative approaches are thus needed that account for the live birth process while addressing interpretable causal questions that are identified under plausible assumptions.

\section{Conditional Stochastic Direct Effects in Perinatal Epidemiology}
 We propose an alternative approach to defining causal parameters in perinatal epidemiology, using direct effects based on stochastic interventions on the mediating variables. Stochastic interventions consider a hypothetical intervention on the mediators where their values are randomly drawn from a distribution (possibly conditional on covariates) rather than deterministically set to a particular value. 
 
Direct effect parameters formulated using stochastic interventions on mediating variables \supercite{didelez_direct_2006, geneletti_identifying_2007,van_der_laan_direct_2008,vanderweele_effect_2014,vanderweele_mediation_2017,zheng_longitudinal_2017} have several advantages, including  
that they define more realistic causal parameters (by allowing for variation in mediator values), reduce positivity concerns, naturally generalize to longitudinal settings, and remain identifiable in the common setting of post-treatment confounding of the mediator-outcome relationship. In perinatal settings, stochastic direct effects are particularly appealing because they can be formulated  to exclude from the total effect any intervention impact that affects infant risk that is mediated by the live birth process. In the remainder of the paper, we discuss the use of conditional stochastic direct effects to frame and identify causal questions in perinatal epidemiology, using the SEARCH Study as a running example.

\subsection{Structural Causal Model}
In many perinatal applications, the observed data consist of baseline covariates, an indicator of exposure, time-varying covariates and mediators, and infant outcome (e.g., HIV-free survival in SEARCH). A longitudinal structural causal model (SCM) can be used to formalize knowledge about the data-generating process.\supercite{pearl_introduction_2010} For simplicity, we focus on an individual-level SCM (noting that this framework generalizes to hierarchical settings\supercite{balzer_new_2019,benitez_defining_2023}). We use $\bar{V}_t$ to denote $(V_1, \dots, V_t)$, which represents the values of the variable $V$ from time $1$ through time $t$. $L_0$ represents  baseline covariates (e.g., maternal age, prior ART use, and community HIV prevalence). $A$ denotes exposure (e.g., if the individual was randomized to receive the intervention). $L_t$ is the set of time-varying covariates (e.g., maternal ART status and infant vital status). $Z_t$ represents the mediator(s) of interest, which in SEARCH consist of $Z1_t$ (maternal mortality by time $t$) and $Z2_t$ (live birth status at time $t$). Let $Y_\tau$ denote the final outcome, where  $\tau$ represents the final time point. In SEARCH, $Y_\tau$ denotes an indicator of an infant being born and surviving HIV-free until the end of follow-up, a composite of $Y1_\tau$ (an indicator of an infant being born and remaining alive) and $Y2_\tau$ (an indicator of an infant remaining HIV-uninfected) by the end of follow-up $\tau$. The exogenous nodes, representing unmeasured variables, are: $U = (U_{L_0}, U_A, U_{L_t}, U_{Z_t}, U_{Y_\tau})$ for $t=1, \dots, \tau$. The endogenous nodes  are: $X = (L_0, A, \bar{L}_t, \bar{Z}_t, Y_\tau)$ for $t=1, \dots, \tau$. In randomized trials such as SEARCH, the exposure assignment mechanism is known. Less is known about the time-varying covariates, mediator and outcome processes, which, in particular, may have been affected by exposure and in turn may have affected subsequent covariates, mediators, and outcomes (Appendix \ref{appendix:scm}).  Figure \ref{dag:interventions}\subref{dag:orig} illustrates (using fewer timepoints) the mechanisms of interest by which the SEARCH intervention may have impacted infant HIV-free survival, which include the intervention's effects through changes to maternal ART initiation, maternal mortality, and live birth.
	
\subsection{Causal Estimands}
The relevance of conditional stochastic direct effect parameters for perinatal epidemiology, as well as the applicability of the CSDE approach for a much broader class of conditional causal effects, is clarified by first revisiting the definition of total effects in these settings. 

\subsubsection{Conditional Total Effects}
Epidemiology commonly studies total effects, defined using marginal summaries of the distribution of counterfactual outcomes. However, in many settings, including perinatal epidemiology, conditional effects defined using summaries of the distribution of counterfactual outcomes \textit{among a counterfactual subset of the population} are often of greater interest. To formulate such conditional total effect parameters, as in Chiu et al.,\supercite{chiu_effect_2020} we use a composite outcome of live birth and infant outcomes. However, in contrast to Chiu et al.,\supercite{chiu_effect_2020} who define the total effect as the effect of the exposure on the probability of the composite event of live birth and infant outcomes, we instead define a conditional total effect (Table \ref{table:sde}): 

	\begin{align*}
	& P[Y1_{\tau}(a=1) = 1, Y2_{\tau}(a=1)=1 | Z2_{\tau}(a=1)=1] \\
	& \quad \quad \quad \quad - P[Y1_{\tau}(a=0) = 1, Y2_{\tau}(a=0)=1 | Z2_{\tau}(a=0)=1]   \\
	&= \frac{P[Y_{\tau}(a=1)=1]}{P[Z2_{\tau}(a=1)=1]} - \frac{P[Y_{\tau}(a=0)=1]}{P[Z2_{\tau}(a=0) =1]} 
	\end{align*}

In the SEARCH Study, this conditional total effect evaluates the effect of the exposure on the outcome of live birth and infant HIV-free survival ($Y1_\tau(a) = 1,Y2_\tau(a) = 1$), but only among counterfactual live births ($Z2_\tau(a) =1$). By the definition of conditional probabilities, the conditional causal effect can alternatively be expressed by contrasting the joint probability of the composite counterfactual outcome of live birth and infant outcomes divided by the probability of counterfactual live birth for each exposure of interest. 

Conditioning on the probability of counterfactual live birth will exclude from the traditional total effect parameter any effect of the intervention on infant outcomes that operates through changes to the marginal probability of live birth by the end of follow-up; however, this conditional total effect parameter does not ensure equivalence of other aspects of the live birth process under differing exposures, and thus will still include any effects due to exposure-induced changes to the the types of people who have a live birth and the timing of births (i.e., composition of the counterfactual population at-risk). For example, in SEARCH, the conditional total effect of the intervention on infant HIV-free survival would include both any direct reduction in  vertical transmission (e.g., by improving HIV viral suppression) and any increase in transmission due to facilitating live birth among persons at higher risk of transmission (e.g., resulting from barriers to care) or changing the time of birth (e.g., such that a higher proportion of follow-up time is prenatal). Such conditional total effects are often of real-world interest in that they reflect the total change in infant outcomes due to multiple mechanisms expected were an intervention to be implemented. Nonetheless, in many cases it remains of interest to isolate intervention effects on outcomes among the same types of people with the same timing of births. 

\subsubsection{Conditional Stochastic Direct Effect}
Conditional stochastic direct effect parameters allow us to quantify the effect of an exposure on infant outcome absent any effect of the intervention on the live birth process, or in other words, the exposure effect in populations with the same types of people (Table \ref{table:sde}). We build on our definition of a conditional total effect to define such a conditional stochastic direct effect. In contrast to a controlled direct effect parameter, which would compare hypothetical interventions in which exposure is varied but all persons in the target population survive to give birth (Figure \ref{dag:interventions}\subref{dag:cde}), stochastic direct effect parameters, \supercite{didelez_direct_2006,geneletti_identifying_2007, van_der_laan_direct_2008} compare hypothetical interventions in which exposure is varied but mediator values $Z_\tau$ are drawn from a distribution that is held constant, denoted here (for ease of of notation) as simply $\bar{G}$. Given some choice of stochastic mediator distribution $\bar{G}$, we use this approach to define the following conditional stochastic direct effect parameter: 
\begin{align*}
	& P[Y1_{\tau}(a=1, \bar{G}) = 1, Y2_{\tau}(a=1, \bar{G})=1 | Z2_{\tau}(\bar{G}) = 1] \\
	 & \quad \quad \quad \quad - P[Y1_{\tau}(a=0, \bar{G}) = 1, Y2_{\tau}(a=0, \bar{G})=1 | Z2_{\tau}(\bar{G}) = 1] \\
	&= \frac{P[Y_{\tau}(a=1, \bar{G}) = 1]}{P(Z2_{\tau}(\bar{G}) = 1)} - \frac{P[Y_{\tau}(a=0, \bar{G}) = 1]}{P(Z2_{\tau}(\bar{G})=1)} \\
	\end{align*}

 As with the CTE, this CSDE is defined among infants born to persons in the target population under each exposure level of interest; however, in contrast to the conditional total effect, the hypothetical intervention to draw maternal survival and live birth from some fixed distribution $\bar{G}$ ensures that the exposure effect is evaluated among the same counterfactual population. For example, in SEARCH, the CSDE corresponds to the difference in counterfactual probability of  HIV-free survival among infants born to persons with HIV under intervention or control, had maternal mortality and live birth over time, $\overline{Z1}_\tau$ and $\overline{Z2}_\tau$, respectively, been drawn from the same underlying distribution. 
 
 In contrast to the principal stratum direct effect, which evaluates the impact of the SEARCH intervention in the group who would give birth regardless of treatment assignment, the stochastic direct effect parameter can evaluate the direct effect of an exposure among different groups of people, depending on the choice of the distribution $\bar{G}$ from which mediator values are drawn. A variety of CSDEs can be formulated using different choices of stochastic interventions on the mediator variables to help understand how the intervention effect varies by the timing of births and composition of people who give birth.

	\afterpage{\begin{landscape}
	\noindent %
	\begin{figure}[H]
    \begin{singlespace}
		\centering
		\begin{subfigure}[t]{0.48\linewidth}
			\caption{}            
			\centering
			\resizebox{0.8\linewidth}{!}{%
			\begin{tikzpicture}

	\node[text centered, text width = 3cm] (W) {\scriptsize Baseline covariates: age, ART naive ($L_0$)};
	
	\node[below = 0.25 of W, text centered, text width = 4cm] (A) {\scriptsize SEARCH intervention ($A$)};

	\node[right = .5 of A, text centered, align = center] (Z2) {\scriptsize Maternal mortality ($Z1$)}; 
	\node[above = .5 of Z2, text centered, align = center] (Z1) {\scriptsize Maternal time-to-ART start ($L_1$)}; 
	\node[below = .5 of Z2, text centered, align = center] (Z3) {\scriptsize Live birth ($Z2$)}; 
	
	\node[right = .5 of Z2, text centered, text width = 2.75cm] (Y) {\scriptsize Live birth by and \\ infant HIV-free survival at end of follow-up \\ ($Y$)};

	 \draw[->, line width= 0.5] (W.20) to [out = 20, in = 120, looseness = 0.5]  (Z1.120);
	 \draw[->, line width= 0.5] (W) --  (Z2);
	 \draw[->, line width= 0.5] (W) --  (Z3);
	 
	 \draw[->, line width= 0.5] (W.80) to [out = 80, in = 90, looseness = 0.5]  (Y.90);

	 \draw[->, line width= 0.5] (A) --  (Z1);
	 \draw[->, line width= 0.5] (A) --  (Z2);
	 \draw[->, line width= 0.5] (A) --  (Z3);
	 
	 \draw[->, line width= 0.5] (A.-80) to [out = -80, in = 230, looseness = 1]  (Y.230);

	 \draw[->, line width= 0.5] (Z1) --  (Z2);
	 \draw[->, line width= 0.5] (Z1.-120) to [out = -120, in = 120, looseness = 0.25]  (Z3.120);
	 
	 \draw[->, line width= 0.5] (Z1) --  (Y);

	 \draw[->, line width= 0.5] (Z2) --  (Z3);
	 
	 \draw[->, line width= 0.5] (Z2) --  (Y);

	 \draw[->, line width= 0.5] (Z3) --  (Y);

\end{tikzpicture}%
			}
			\label{dag:orig}
		\end{subfigure}
		\hfill
		\begin{subfigure}[t]{0.48\linewidth}
			\caption{}  
			\centering
			\resizebox{\linewidth}{!}{%
			\begin{tikzpicture}

	\node[text centered, text width = 3cm] (W) {\scriptsize Baseline covariates: age, ART naive ($L_0$)};
	
	\node[below = 0.25 of W, text centered, text width = 4cm] (A) {\scriptsize SEARCH intervention ($A$)};

	\node[right = .5 of A, text centered, align = center] (Z2) {\scriptsize Maternal mortality ($Z1$)}; 
	\node[above = .5 of Z2, text centered, align = center] (Z1) {\scriptsize Maternal time-to-ART start ($L_1$)}; 
	\node[below = .5 of Z2, text centered, align = center] (Z3) {\scriptsize Live birth ($Z2$)}; 
	
	\node[right = .5 of Z2, text centered, text width = 2.75cm] (Y) {\scriptsize Live birth by and \\ infant HIV-free survival at end of follow-up \\ ($Y$)};

	 \draw[->, line width= 0.5] (W.20) to [out = 20, in = 120, looseness = 0.5]  (Z1.120);
	 
	 \draw[->, line width= 0.5] (W.80) to [out = 80, in = 90, looseness = 0.5]  (Y.90);

	 \draw[->, line width= 0.5] (A) --  (Z1);
	 
	 \draw[->, line width= 0.5] (A.-80) to [out = -80, in = 230, looseness = 1]  (Y.230);

	 \draw[->, line width= 0.5] (Z1) --  (Y);

	 \draw[->, line width= 0.5] (Z2) --  (Z3);
	 
	 \draw[->, line width= 0.5] (Z2) --  (Y);

	 \draw[->, line width= 0.5] (Z3) --  (Y);

	\node[above left = 0.25 of A, text centered] (a) {\scriptsize $A = a \text{ for } a \in \{0,1\}$}; 
	\node[below left = 2 of Z2, text centered] (g) {\scriptsize $Z1 = 0, Z2 = 1$};
	
		\draw[->, line width = 0.5, cyan] (a) -- (A.170);	 

		\draw[->, line width = 0.5, cyan] (g) -- (Z2.-160);
		\draw[->, line width = 0.5, cyan] (g) -- (Z3);

\end{tikzpicture}%
			}
			\label{dag:cde}
		\end{subfigure}
		\hfill
		\begin{subfigure}[t]{0.48\linewidth}
 			\caption{} 
			\centering
			\resizebox{\linewidth}{!}{%
			\begin{tikzpicture}

	\node[text centered, text width = 3cm] (W) {\scriptsize Baseline covariates: age, ART naive ($L_0$)};
	
	\node[below = 0.25 of W, text centered, text width = 4cm] (A) {\scriptsize SEARCH intervention ($A$)};

	\node[right = .5 of A, text centered, align = center] (Z2) {\scriptsize Maternal mortality ($Z1$)}; 
	\node[above = .5 of Z2, text centered, align = center] (Z1) {\scriptsize Maternal time-to-ART start ($L_1$)}; 
	\node[below = .5 of Z2, text centered, align = center] (Z3) {\scriptsize Live birth ($Z2$)}; 
	
	\node[right = .5 of Z2, text centered, text width = 2.75cm] (Y) {\scriptsize Live birth by and \\ infant HIV-free survival at end of follow-up \\ ($Y$)};

	 \draw[->, line width= 0.5] (W.20) to [out = 20, in = 120, looseness = 0.5]  (Z1.120);
	 \draw[->, line width= 0.5, cyan] (W) --  (Z2);
	 \draw[->, line width= 0.5, cyan] (W) --  (Z3);
	 
	 \draw[->, line width= 0.5] (W.80) to [out = 80, in = 90, looseness = 0.5]  (Y.90);

	 \draw[->, line width= 0.5] (A) --  (Z1);

	 \draw[->, line width= 0.5] (A.-80) to [out = -80, in = 230, looseness = 1]  (Y.230);

	 \draw[->, line width= 0.5] (Z1) --  (Y);

	 \draw[->, line width= 0.5] (Z2) --  (Z3);
	 
	 \draw[->, line width= 0.5] (Z2) --  (Y);

	 \draw[->, line width= 0.5] (Z3) --  (Y);

	\node[above left = 0.25 of A, text centered] (a) {\scriptsize $A = a \text{ for } a \in \{0,1\}$};

		\draw[->, line width = 0.5, cyan] (a) -- (A.170);	 

\end{tikzpicture}%
			}
			\label{dag:sde}
		\end{subfigure}
		\hfill
		\begin{subfigure}[t]{0.48\linewidth}
  			\caption{}
			\centering
			\resizebox{\linewidth}{!}{%
			\begin{tikzpicture}

	\node[text centered, text width = 3cm] (W) {\scriptsize Baseline covariates: age, ART naive ($L_0$)};
	
	\node[below = 0.25 of W, text centered, text width = 4cm] (A) {\scriptsize SEARCH intervention ($A$)};

	\node[right = .5 of A, text centered, align = center] (Z2) {\scriptsize Maternal mortality ($Z1$)}; 
	\node[above = .5 of Z2, text centered, align = center] (Z1) {\scriptsize Maternal time-to-ART start ($L_1$)}; 
	\node[below = .5 of Z2, text centered, align = center] (Z3) {\scriptsize Live birth ($Z2$)}; 
	
	\node[right = .5 of Z2, text centered, text width = 2.75cm] (Y) {\scriptsize Live birth by and \\ infant HIV-free survival at end of follow-up \\ ($Y$)};

	 \draw[->, line width= 0.5] (W.20) to [out = 20, in = 120, looseness = 0.5]  (Z1.120);
	 \draw[->, line width= 0.5, cyan] (W) --  (Z2);
	 \draw[->, line width= 0.5, cyan] (W) --  (Z3);
	 
	 \draw[->, line width= 0.5] (W.80) to [out = 80, in = 90, looseness = 0.5]  (Y.90);

	 \draw[->, line width= 0.5] (A) --  (Z1);
	 
	 \draw[->, line width= 0.5] (A.-80) to [out = -80, in = 230, looseness = 1]  (Y.230);

	 \draw[->, line width= 0.5, cyan] (Z1) --  (Z2);
	 \draw[->, line width= 0.5, cyan] (Z1.-120) to [out = -120, in = 120, looseness = 0.25]  (Z3.120);
	 
	 \draw[->, line width= 0.5] (Z1) --  (Y);

	 \draw[->, line width= 0.5, cyan] (Z2) --  (Z3);
	 
	 \draw[->, line width= 0.5] (Z2) --  (Y);

	 \draw[->, line width= 0.5] (Z3) --  (Y);

	\node[above left = 0.25 of A, text centered] (a) {\scriptsize $A = a \text{ for } a \in \{0,1\}$};

		\draw[->, line width = 0.5, cyan] (a) -- (A.170);	 

\end{tikzpicture}%
			}	
			\label{dag:nde}
		\end{subfigure}
		
	\caption{(A) A simplified DAG with post-treatment covariates and mediators at one time point in the SEARCH Study. (B) A post-intervention DAG illustrating a controlled direct effect parameter which enforces maternal survival and live birth. (C) A post-intervention DAG illustrating a stochastic direct effect parameter where in the hypothetical intervention, mediators are drawn conditional only on baseline covariates. (D) A post-intervention DAG representing a stochastic direct effect parameter where the hypothetical intervention on the live birth process is drawn conditional on the baseline covariates and time-varying past.}
	\label{dag:interventions}
    \end{singlespace}
	\end{figure}
	\end{landscape}}

\subsubsection{Choice of Mediator Distribution: Interpretation and Identification}
In specifying a distribution $\bar{G}$ from which the mediator value(s) are drawn, there are several options, with implications for both parameter identification (or ability to rewrite the causal parameter as a parameter of the observed data distribution and thus estimate it) and interpretation. One approach\supercite{van_der_laan_direct_2005, van_der_laan_direct_2008, didelez_direct_2006} is to define the (potentially longitudinal) mediator distribution as the unknown counterfactual distribution in the absence of exposure, possibly conditional on baseline covariates (also referred to as a marginal mediator or fixed stochastic direct effect in the context of non-conditional effect \supercite{vanderweele_mediation_2017, rudolph_robust_2017} (Appendix \ref{appendix:nde})). In perinatal settings, such a natural direct effect formulated conditional on baseline covariates holds constant the types and timing of births (Figure \ref{dag:interventions}\subref{dag:sde}) across exposure conditions; however, in the presence of time-varying confounding, it will not decompose the original total effect. \supercite{zheng_longitudinal_2017, vanderweele_mediation_2017}  
Identification of this parameter requires no unmeasured confounding of the 1) exposure-outcome, 2) exposure-mediator, and 3) mediator-outcome relationships at all time points (but  avoids a stronger "cross-world" assumption required by earlier definitions\supercite{robins_identifiability_1992, pearl_direct_2001}). Identification requires a positive probability of 1) the values the mediator takes under the control condition given each exposure of interest within all values of the mediator-outcome confounders (baseline and time-varying covariates) at each time $t$ and 2) of each possible exposure within all values of baseline covariates.

Another choice for the stochastic intervention on the mediator(s) is to draw mediator values from the counterfactual control distribution conditional both on baseline covariates and time-varying history, as proposed by Zheng and van der Laan.\supercite{zheng_longitudinal_2017} While this conditional mediator formulation of the natural direct effect is appealing because it provides a decomposition of the average treatment effect,\supercite{zheng_longitudinal_2017} in perinatal settings, this type of parameter does not ensure that the timing of births and types of people who give birth under different exposures will be the same (i.e., does not ensure equivalent counterfactual populations at-risk across differing exposures). In perinatal settings (Figure \ref{dag:interventions}\subref{dag:nde}), this corresponds to a hypothetical intervention where live birth is a function of both the baseline covariates and the time-varying past (e.g., maternal time-to-ART start). Because the time-varying past is affected by the static intervention on $A$, the types of people who have a live birth under an intervention to assign everyone to receive the intervention may not be the same as those who would have a live birth if everyone were to be assigned to the control condition. This parameter also requires stronger assumptions for identifiability than the natural direct effect using marginal mediator distributions (i.e., strong sequential randomization and its associated positivity assumptions).\supercite{zheng_longitudinal_2017} 

Alternatively, conditional stochastic direct effect parameters can be formulated such that mediator values are drawn from some known distribution, conditional only on baseline covariates (Figure \ref{dag:interventions}\subref{dag:sde}), or from a corresponding distribution that is estimated from the observed data (i.e., a data-adaptive parameter as proposed by Rudolph et al.\supercite{hubbard_statistical_2016, rudolph_robust_2017}). Data-dependent choices for the stochastic intervention on the mediator include the estimated mediator distribution in the intervention or comparator conditions or an intervention that is well-supported in the study sample. In SEARCH, for example, we could formulate a data-adaptive conditional stochastic direct effect parameter which compared infant HIV-free survival among live births if everyone received the intervention versus the control condition and if maternal mortality and time-to-live birth were drawn from their observed distribution in the intervention arm conditional on baseline covariates, such as maternal age, community-specific HIV prevalence, and baseline ART initiation status. 

Such stochastic direct effect parameters, in which the distribution from which the mediators is drawn is known or estimated, require fewer assumptions for identifiability than mediated effects formulated using either the marginal or conditional mediator approaches described previously. More specifically, identification requires no unmeasured confounding of the 1) exposure-outcome and 2) mediator-outcome relationships at all time points. In addition, identification requires positive probabilities of 1) each exposure of interest given baseline covariates and 2) of the values the mediator can take under its specified known distribution given each exposure within all past values of mediator-outcome confounders (baseline and time-varying factors) at each timepoint. If the observed mediator distribution under one of the exposure values is selected as the known distribution from which mediator values are drawn, then the positivity assumption will be satisfied for that value of the exposure variable. 

The assumption of no unmeasured confounding of the exposure and outcome will be satisfied by design in a randomized controlled trial such as the SEARCH Study. Randomization of the exposure variable is not sufficient to satisfy the second assumption of no unmeasured confounding of the mediator-outcome relationship(s). In the SEARCH trial, the second assumption will be satisfied if we believe that all confounders of the live birth process (maternal mortality and live birth at time $t$) and outcome (infant HIV-free survival) at each time point were measured. In SEARCH, a potential set of such variables includes maternal ART initiation through time $t$, mediator history (maternal mortality and live birth through time $t-1$), and baseline covariates. Measurement of additional time-varying covariates, such as maternal health (e.g., CD4 cell count and viral load) over time, makes it more likely that this assumption will be satisfied. In general, measuring a rich set of baseline and time-varying covariates will improve the plausibility of both of these assumptions on confounding.

The conditional total effect will be identified if there is 1) no unmeasured exposure-outcome confounding and 2) a positive probability of each exposure within all baseline covariate values. Because the outcome of interest is defined as the composite event of the mediator live birth and infant outcomes, in SEARCH, this randomization assumption more specifically corresponds to assumptions of no unmeasured confounding of the exposure-live birth ($A-Z2_{\tau}$), exposure-infant HIV status ($A-Y1_{\tau}$), and exposure-infant vital status ($A-Y2_{\tau}$) relationships. Unlike the CTE, the CSDE does not require an assumption of no unmeasured confounding of the exposure-live birth relationship because the mediator, the live birth process ($\bar{Z}_{\tau}$), is intervened upon in the formulation of the CSDE parameter.\supercite{pearl_probabilistic_1995} As a result, in observational settings, it is possible that the CSDE may be identified even when the CTE is not, due to unmeasured exposure-mediator confounding. (In the setting of a randomized trial, randomization assumptions involving the exposure will be satisfied by design.)

	\begin{sidewaystable}[ph!]
	\centering
    \caption{Definition and interpretations of causal estimands proposed for the perinatal epidemiologic setting.}
	\scriptsize
	\begin{tabular}{R{6.8cm}R{3.4cm}R{3.1cm}R{3.4cm}R{3.4cm}} \toprule
      Parameter &  Interpretation (General) & Interpretation (SEARCH) & Counterfactuals (SEARCH) & Hypothetical Intervention\\ \midrule
        \label{table:sde}
	\textbf{Total effect parameter} & & & & \\ 
	$\begin{aligned}[t]
	& \frac{P[Y_{\tau}(a=1)=1]}{P[Z2_{\tau}(a=1)=1]} - \frac{P[Y_{\tau}(a=0)=1]}{P[Z2_{\tau}(a=0) =1]} \\
	&= P[Y1_{\tau}(a=1) = 0, Y2_{\tau}(a=1)=0 | Z2_{\tau}(a=1)=1]  \\
	&- P[Y1_{\tau}(a=0) = 0, Y2_{\tau}(a=0)=0 | Z2_{\tau}(a=0)=1]   
	\end{aligned}$ 
		& \vspace{-1em}\begin{itemize}[leftmargin=*, noitemsep, topsep=0pt]
			\item The total effect of an intervention on infant health outcomes among live births
			\item Includes: any effect of the intervention operating through changes to the timing and types of people who give birth 
			\item Excludes: any effect of the intervention on infant health outcomes through changes to the marginal probability of live birth by the end of the follow-up period
			\end{itemize} 
		& \vspace{-1em}\begin{itemize}[leftmargin=*, noitemsep, topsep=0pt]
			\item The total effect of the SEARCH intervention on infant HIV-free survival among live births
			\end{itemize}  
		& \vspace{-1em}\begin{itemize}[leftmargin=*, noitemsep, topsep=0pt]
			\item $Y_{\tau}(a)$: counterfactual outcome of live birth and infant HIV-free survival at time $\tau$ when $A=a$
			\item $Y1_{\tau}(a)$: counterfactual infant vital status at time $\tau$ when $A=a$
			\item $Y2_{\tau}(a)$: counterfactual infant HIV status at time $\tau$ when $A=a$
			\item $Z2_{\tau}(a)$: counterfactual live birth status by time $\tau$ when $A=a$
			\end{itemize}
		& \vspace{-1em}\begin{itemize}[leftmargin=*, noitemsep, topsep=0pt]
			\item Compare intervening to set $a=1$ (i.e., receive the SEARCH intervention) for everyone to setting $a=0$ (i.e., receive country standard of care) for everyone 
			\end{itemize} \\
	\textbf{Stochastic direct effect parameters} & & & & \\ 
	$\begin{aligned}[t]
	& \frac{P[Y_{\tau}(a=1, \bar{G}) = 1]}{P(Z2_{\tau}(\bar{G}) = 1)} - \frac{P[Y_{\tau}(a=0, \bar{G}) = 1]}{P(Z2_{\tau}(\bar{G})=1)} \\
	&= P[Y1_{\tau}(a=1, \bar{G}) = 0, Y2_{\tau}(a=1, \bar{G})=0 | Z2_{\tau}(\bar{G}) = 1] \\
	&- P[Y1_{\tau}(a=0, \bar{G}) = 0, Y2_{\tau}(a=0, \bar{G})=0 | Z2_{\tau}(\bar{G}) = 1] 
	\end{aligned}$ \newline \vspace{1em} where $\bar{G} \equiv \bar{G}_{\tau}$ and $\bar{G}_t\big(\bar{z}_t | L_0\big) \equiv P\big(\bar{Z}_t = \bar{z}_t | L_0\big)$ is a known distribution (conditional on baseline covariates)\tablefootnote{In the specific case where the known distribution of the mediators conditional on baseline covariates is equivalent to the true marginal mediator distribution conditional on baseline covariates in the control arm, this stochastic direct effect parameter will be equal to the natural direct effect parameter formulated using marginal mediator distributions (and will provide a decomposition of a type of total effect).}
		& \vspace{-1em}\begin{itemize}[leftmargin=*, noitemsep, topsep=0pt]
			\item A direct effect of an intervention on infant health outcomes where the live birth process (a person's probability of giving birth at each time point) is drawn from the same known distribution conditional on baseline covariates
			\item Includes: any effect of the intervention on the outcome that operates through mechanisms other than the live birth process
			\item Excludes: any effect of the intervention on infant health outcomes through changes to the mediator: timing/occurrence of live births and to the types of people who give birth
			\end{itemize} 
		& \vspace{-1em}\begin{itemize}[leftmargin=*, noitemsep, topsep=0pt]
			\item The direct effect of the SEARCH intervention on infant HIV-free survival if the live birth process was drawn from some known distribution conditional on the baseline covariates of a woman in the target population
			\end{itemize} 
		& \vspace{-1em}\begin{itemize}[leftmargin=*, noitemsep, topsep=0pt] 
			\item $Y_{\tau}(a, \bar{G})$: counterfactual outcome of live birth and infant HIV-free survival at $t=\tau$ under an intervention to set $A=a$ and draw maternal survival and time-to-live birth from a known distribution that is conditional on observed baseline covariates
			\item $Y1_{\tau}(a', \bar{G})$: counterfactual infant vital status at $t=\tau$ under an intervention to set $A=a$ and draw maternal survival and time-to-live birth from the known mediator distribution 
			\item $Y2_{\tau}(a', \bar{G})$: counterfactual infant HIV status at $t=\tau$ under an intervention to set $A=a$ and draw maternal survival and time-to-live birth from the known mediator distribution 			
			\item $Z2_{\tau}(\bar{G})$: counterfactual live birth status at $t=\tau$ under the known stochastic intervention on the mediators
			\end{itemize}
		& \vspace{-1em}\begin{itemize}[leftmargin=*, noitemsep, topsep=0pt]
			\item Compare intervening to set $a=1$ to setting $a=0$ for all persons in the target population, while drawing the live birth process from the same known distribution conditional on baseline covariates
			\end{itemize} \\\\		
	\bottomrule
    \end{tabular}
    \end{sidewaystable}

\section{Discussion}
A common challenge of causal effect estimation in perinatal epidemiology is that outcomes are only measured among infants born alive, but exposures typically predate birth and may affect the live birth process, resulting in different types of people who have live births. While particularly notable in the perinatal epidemiology literature, this challenge also occurs ubiquitously across settings in  where an exposure may affect the composition of the group that is at risk for the study outcome.\supercite{balzer_far_2020} We add to existing literature addressing this challenge by first explicitly defining and interpreting conditional total effects in this setting, highlighting two components of the total effect, one due to due to exposure effects among counterfactual group members, and a second due to exposure effects on counterfactual group membership. 

We then suggest the use of a class of conditional stochastic direct effects to disentangle these two exposure pathways and discuss setting-specific advantages over other types of direct effect parameters described in the perinatal literature. Such conditional direct effect parameters are formulated using stochastic interventions on the mediator process conditional on baseline covariates. The choice of stochastic intervention on the mediator(s) requires careful consideration as it determines the subgroup in which the intervention is assessed. In perinatal settings, the choice of stochastic intervention will determine the timing of births and composition of people who have live births, and will thus affect the interpretation of the stochastic direct effect parameter. This flexibility also represents an advantage of this approach; applying different choices of stochastic interventions on the mediating variables can reveal how the direct effect of the intervention is mediated by the live birth process that occurs in perinatal studies, as well as providing a straightforward approach to transporting (under assumptions) perinatal exposure effect estimates to new settings with differing live birth processes, or more generally, of transporting stochastic direct effects among selected subgroups to new settings in which the group selection process my differ.  

Approaches for the estimation of stochastic direct effect parameters have been developed by van der Laan and Petersen,\supercite{van_der_laan_direct_2008} Rudolph et al.,\supercite{rudolph_robust_2017} Zheng and van der Laan,\supercite{zheng_longitudinal_2017} VanderWeele and Tchetgen Tchetgen,\supercite{vanderweele_mediation_2017} and Diaz et al.,\supercite{diaz_nonparametric_2021} among others, and include a range of inverse probability weighted and semiparametric efficient multiply robust estimators. However, to date, while stochastic mediation effects have been applied in perinatal studies previously,\supercite{vanderweele_effect_2014, goin_mediation_2020} to the best of our knowledge, they have not been used to formulate conditional causal effect parameters that account for exposure-induced changes to the live birth process in these types of settings, nor more generally to account for exposure-induced changes in subgroup membership when conditional causal effects are of interest. This paper aims to provide a bridge to the wider use of these methods in settings where they are perhaps uniquely positioned to address the underlying question of interest. 

\section*{Acknowledgements}
\noindent We thank the SEARCH Collaboration, Patrick Bradshaw, the Ministry of Health of Uganda and the Ministry of Health of Kenya, our research teams and administrative teams in San Francisco, Uganda, and Kenya, collaborators and advisory boards, and especially all the communities and participants involved in the SEARCH trial.

\newpage
\singlespacing
\printbibliography

\newpage
\begin{appendices}
\doublespacing
\section{Structural Causal Model}\label{appendix:scm}

A longitudinal structural causal model is used to represent the data generating process of perinatal HIV transmission in the target population of women living with HIV who are at risk for pregnancy over the follow-up period. $L_0$ denotes the baseline covariates and $A$ the intervention node. $L_t$ represents time-varying covariates (which may be confounders of the mediator-outcome relationship(s)) at each time $t$. $Z_t$ denotes the set of mediators, $Z1_t$ and $Z2_t$, at each time point. We use two mediators to capture two mechanisms by which the intervention may affect the timing of births and the types of people who give birth: 1) by the effect of the intervention on maternal morality ($Z1_t$) and 2) by the effect of the intervention on the timing of live birth ($Z2_t$). Both $Z1_t$ and $Z2_t$ are time-to-event variables that each take a value of 1 deterministically at time $t+1$ if they are equal to 1 at time $t$. The final outcome of interest is the composite variable $Y_{\tau}$, which takes a value of 1 if, in terms of our motivating example, a live birth occurs and the infant survives and is HIV-free at the end of follow-up and a value of 0 otherwise. Infant survival is represented by the node $Y1_t$; its value is missing if a live birth has not yet occurred by time $t$. $Y2_{\tau}$ denotes whether a live infant was HIV-free at the end of follow-up. Because infection status was only assessed at the end of the study, it is not a time-varying covariate, and its value will be missing in the absence of a live birth or in the event of infant mortality prior to the end of the follow-up period.

\begin{align*}
       L_0 &= f_{L_0}(U_{L_0}) \\
       A &= f_A(U_A) \\
       L_t &= f_{L_t}(A, \bar{L}_{t-1}, \bar{Z}_{t-1}, U_{L_t}) \\
       Z_t &= \{Z1_t, Z2_t\} \\
       Z1_t &= 
       		\begin{cases}
		1, & \text{ if } Z1_{t-1} = 1 \\
		f_{Z1_t}(A, \bar{L}_t, \bar{Z}_{t-1}, U_{Z_t}), &\text{ otherwise } \\
		\end{cases} \\
	Z2_t &= 
		\begin{cases}
		1, & \text{ if } Z2_{t-1} = 1 \\
		f_{Z2_t}(A, \bar{L}_t, \bar{Z}_{t-1}, U_{Z_t}), &\text{ otherwise } \\
		\end{cases} \\	
	Y1_t &\in L_{t+1}\\
	Y1_t &=
         	\begin{cases}
            	NA, & \text{ if } Z2_t = 0 \\
            	0, & \text{ if } Y1_{t-1} = 0 \\
            	f_{Y1_t}(A, \bar{L}_t, \bar{Z}_t, U_{Y1_t}), & \text{ otherwise } \\
            	\end{cases} \\
	Y2_{\tau} &= 
    		\begin{cases}
            	NA, & \text{ if } Z2_{\tau} = 0 \text{ or } Y1_{\tau} = 0 \\ 
            	f_{Y2_{\tau}}(A, \bar{L}_{\tau}, \bar{Z}_{\tau}, U_{Y2_{\tau}}), & \text{ otherwise } \\
            	\end{cases} \\
	Y_{\tau} &= 
    		\begin{cases}
    		1, & \text{ if } Z2_{\tau} = 1, Y1_{\tau} = 1, \text{ and } Y2_{\tau} = 1 \\
    		0, & \text{ if } Z2_{\tau} = 0 \text{ or } (Z2_{\tau} = 1 \text{ and } (Y1_{\tau} = 0 \text{ or } Y2_{\tau} = 0)) \\
    		\end{cases}
	\end{align*}
	
\section{Summary of Natural Direct Effect Parameters}\label{appendix:nde}

	\begin{sidewaystable}[ph!]
	\centering
    \caption{Definition and interpretations of alternative target natural direct effect parameters.} 
	\scriptsize
	\begin{tabular}{R{8.3cm}R{4cm}R{2.5cm}R{6cm}R{2.7cm}} \toprule
	Parameter & Interpretation (General) & Interpretation (SEARCH) & Counterfactuals (SEARCH) & Hypothetical Intervention \\ \midrule
 	\label{table:nde}
	\textbf{Marginal mediator distributions}\tablefootnote{Denominator is equal to $P[Z2_\tau(a=0)=1]$} & & & & \\ 
	$\begin{aligned}[t]
	& \frac{P[Y_{\tau}(a=1, \bar{G}^{a=0}) = 1]}{P(Z2_{\tau}(\bar{G}^{a=0}) = 1)} - \frac{P[Y_{\tau}(a=0, \bar{G}^{a=0}) = 1]}{P(Z2_{\tau}(\bar{G}^{a=0})=1)} \\
	&= P[Y1_{\tau}(a=1, \bar{G}^{a=0}) = 0, Y2_{\tau}(a=1, \bar{G}^{a=0}) = 0 | Z2_{\tau}(\bar{G}^{a=0}) = 1] \\
	&= P[Y1_{\tau}(a=0, \bar{G}^{a=0}) = 0, Y2_{\tau}(a=0, \bar{G}^{a=0}) = 0 | Z2_{\tau}(\bar{G}^{a=0}) = 1] 
	\end{aligned}$ \newline \vspace{1em} where $\bar{G}^{a=0} \equiv (G_1^{a=0}, \dots, G_{\tau}^{a=0})$ and \newline ${G}_t^{a=0}\big(z_t | l_0\big) \equiv P\big(Z_t(a=0) = z_t | L_0 = l_0\big)$
		& \vspace{-1em}\begin{itemize}[leftmargin=*, noitemsep, topsep=0pt]
			\item A natural direct effect of an intervention on infant health outcomes where the live birth process is drawn from its true distribution of live births conditional on baseline covariates under the control condition
			\item Includes: any effect of the intervention on the outcome that operates through mechanisms other than the live birth process
			\item Excludes: any effect of the intervention on infant health outcomes through changes to the timing/occurrence of live births and to the types of people who give birth
			\end{itemize} 
		& \vspace{-1em}\begin{itemize}[leftmargin=*, noitemsep, topsep=0pt]
			\item The natural direct effect of the SEARCH intervention on infant health outcomes where the live birth process is drawn from the true distribution of live births conditional on baseline covariates under the control condition
			\end{itemize}
		& \vspace{-1em}\begin{itemize}[leftmargin=*, noitemsep, topsep=0pt]
			\item $Y_{\tau}(a, \bar{G}^{a'})$ represents counterfactual live birth and infant HIV-free survival at $t=\tau$ under an intervention to set $A=a$ and draw the mediators from their (unknown) distribution under $A=a'$ conditional on baseline covariates
			\item $Y1_{\tau}(a, \bar{G}^{a'})$ represents counterfactual infant vital status at $t=\tau$ under an intervention to set $A=a'$ and draw maternal survival and time-to-live birth from their distribution under $A=a'$ conditional on baseline covariates
			\item $Y2_{\tau}(a, \bar{G}^{a'})$ represents counterfactual infant HIV status at $t=\tau$ under an intervention to set $A=a'$ and draw maternal survival and time-to-live birth from their distribution under $A=a'$ conditional on baseline covariates
			\item $Z2_{\tau}(\bar{G}^{a'})$ is the counterfactual live birth status at $t=\tau$ under the stochastic intervention			
			\end{itemize}
		& \vspace{-1em}\begin{itemize}[leftmargin=*, noitemsep, topsep=0pt]
			\item Compare intervening to set $A=1$ for everyone to setting $A=0$ for everyone and drawing the live birth process from the true, unknown live birth distribution conditional on baseline covariates under the control exposure
			\end{itemize}  \\
	\textbf{Conditional mediator distributions} & & & & \\ 
	$\begin{aligned}[t]
	& \frac{P[Y_{\tau}(a=1, \bar{\Gamma}^{a=0}) = 1]}{P(Z2_{\tau}(\bar{\Gamma}^{a=0}) = 1)} - \frac{P[Y_{\tau}(a=0) = 1]}{P(Z2_{\tau}(a=0)=1)} \\
	&= P[Y1_{\tau}(a=1, \bar{\Gamma}^{a=0}) = 0, Y2_{\tau}(a=1, \bar{\Gamma}^{a=0}) = 0 | Z2_{\tau}(\bar{\Gamma}^{a=0}) = 1] \\
	&-  P[Y1_{\tau}(a=0) = 0, Y2_{\tau}(a=0) = 0 | Z2_{\tau}(a=0) = 1] 
	\end{aligned}$ \newline \vspace{1em} where $\bar{\Gamma}^{a=0} \equiv (\Gamma_1^{a=0}, \dots, \Gamma_{\tau}^{a=0})$ and \newline ${\Gamma}_t^{a=0}\big(z_t | \bar{l}_t, \bar{z}_{t-1}\big) \equiv P\big(Z_t(a=0) = z_t | \bar{L}_t(a=0) = \bar{l}_t, \bar{Z}_{t-1}(a=0) = \bar{z}_{t-1}\big)$
		& \vspace{-1em}\begin{itemize}[leftmargin=*, noitemsep, topsep=0pt]
			\item A natural direct effect of an intervention on infant health outcomes where the live birth process is drawn from its true distribution of live births conditional on baseline and time-varying history under the control condition
			\item Includes: any effect of the intervention on the outcome through changes to time-varying covariates which may in turn impact the timing/occurrence of live births and the types of people who give birth
			\item Excludes: any effect of the intervention on infant health outcomes that operating only through the survival of those at risk for pregnancy and/or live birth 
			\end{itemize} 
		& \vspace{-1em}\begin{itemize}[leftmargin=*, noitemsep, topsep=0pt]
			\item The natural direct effect of the SEARCH intervention on infant HIV-free survival if the live birth process was drawn from its true, unknown distribution conditional on baseline and time-varying covariate history under the control condition
			\end{itemize} 
		& \vspace{-1em}\begin{itemize}[leftmargin=*, noitemsep, topsep=0pt] 
			\item $Y_{\tau}(a, \bar{\Gamma}^{a'})$ is counterfactual live birth and infant HIV-free survival at $t=\tau$ under an intervention to set $A=a$ and draw maternal survival and time-to-live birth from their distribution under $A=a'$ conditional on observed baseline and time-varying covariate history
			\item $Y_{\tau}(a)$ is counterfactual live birth and infant HIV-free survival at $t=\tau$ under an intervention to set $A=a$ 
			\item $Y1_{\tau}(a, \bar{\Gamma}^{a'})$ is counterfactual infant vital status at $t=\tau$ under an intervention to set $A=a$ and draw maternal survival and time-to-live birth from their distribution under $A=a'$ conditional on observed baseline and time-varying covariate history
			\item $Y1_{\tau}(a)$ is counterfactual infant vital status at $t=\tau$ under an intervention to set $A=a$ 				\item $Y2_{\tau}(a, \bar{\Gamma}^{a'})$ is counterfactual infant HIV status at $t=\tau$ under an intervention to set $A=a$ and draw maternal survival and time-to-live birth from their distribution under $A=a'$ conditional on observed baseline and time-varying covariate history
			\item $Y2_{\tau}(a)$ is counterfactual infant HIV status at $t=\tau$ under an intervention to set $A=a$ 
			\item $Z2_{\tau}(a, \bar{\Gamma}^{a'})$ is the counterfactual live birth status at $t=\tau$ under this stochastic intervention conditional on baseline and time-varying covariate history
			\item $Z2_{\tau}(a)$ is the counterfactual live birth at $t=\tau$ under an intervention to set $A=a$			\end{itemize}
		& \vspace{-1em}\begin{itemize}[leftmargin=*, noitemsep, topsep=0pt]
			\item Compare intervening to set $A=1$ for everyone to setting $A=0$ for everyone while drawing the live birth process from its true, unknown distribution conditional on baseline and time-varying history 
			\end{itemize} \\		
	\bottomrule
	\end{tabular}
    \end{sidewaystable}

\end{appendices}

\end{document}